\documentclass[useAMS,usenatbib]{mn2e}
\usepackage{amsmath,astrojournals,fleqn,epsfig,txfonts}

\newcommand {\B}[1]{{\boldsymbol{#1}}}
\newcommand {\vsun}{{\B{\upsilon_\odot}}}
\newcommand {\Usun}{{U_{\!\odot}}}
\newcommand {\Wsun}{{W_{\!\odot}}}
\newcommand {\Vsun}{{V_{\!\odot}}}
\newcommand {\BV}{\mbox{$B\,{-}\,V$}}
\newcommand {\figref}[1]{\mbox{Fig.\,\ref{#1}}}
\newcommand {\kms}{\mbox{\,km\,s$^{-1}$}}
\newcommand {\vrq}{\overline{\upsilon^2_{\!R}}}
\newcommand {\va}{\upsilon_{\mathrm{a}}}
\newcommand {\vc}{\upsilon_{\mathrm{c}}}

\newcommand {\Gyr}{\,{\rm Gyr}}
\title{Local Kinematics and the Local Standard of Rest}
\author[R. Sch\"onrich, J. Binney \& W. Dehnen]
       {Ralph Sch\"onrich$^1$\thanks{E-mail: rasch@mpa-garching.mpg.de},
        James Binney$^2$ and Walter Dehnen$^3$\\
        $^{1}$ Max-Planck-Institut f\"ur Astrophysik, Karl-Schwarzschild-Str.~1,
        85741 Garching, Germany \\ 
        $^{2}$ Rudolf Peierls Centre for Theoretical Physics, Keble Road,
        Oxford OX1 3NP, UK\\
        $^{3}$ Department for Physics \& Astronomy, University of Leicester,
        University Road, Leicester LE1 7RH, UK}
\date{Draft, \today}
\pagerange{\pageref{firstpage}--\pageref{lastpage}}
\pubyear{2009}

\begin{document}
\maketitle
\label{firstpage}
\begin{abstract}
  We re-examine the stellar kinematics of the Solar neighbourhood  in terms of the
  velocity $\vsun$ of the Sun with respect to the local standard of rest. We
  show that the classical determination of its component $\Vsun$ in the direction
  of Galactic rotation via Str\"omberg's relation is undermined by the
  metallicity gradient in the disc, which introduces a correlation between the
  colour of a group of stars and the radial gradients of its properties.
  Comparing the local stellar kinematics to a chemodynamical model which
  accounts for these effects, we obtain $(U,V,W)_\odot =
  (11.1_{-0.75}^{+0.69},12.24_{-0.47}^{+0.47},7.25_{-0.36}^{+0.37})\kms$,
  with additional systematic uncertainties $\sim(1,2,0.5)\kms$. In
  particular, $\Vsun$ is $7\kms$ larger than previously estimated. The new
  values of $(U,V,W)_\odot$ are extremely insensitive to the metallicity
  gradient within the disc.
\end{abstract}


\section{Introduction}
The Sun's velocity $\vsun$ with respect to the Local Standard of Rest
(LSR)\footnote{The LSR is the rest frame at the location of the Sun of a star
  that would be on a circular orbit in the gravitational potential one would
  obtain by azimuthally averaging away non-axisymmetric features in the actual
  Galactic potential.} is required to transform any observed heliocentric
velocity to a local galactic frame. Since this transformation is often
necessary for scientific interpretation of observed velocities in terms of
Galactic structure, the determination of $\vsun$ is a fundamental task of
Galactic astronomy. The radial and vertical components $\Usun$ and $\Wsun$ of
$\vsun$ are straightforwardly obtained from the mean heliocentric velocities
of several different groups of Solar-neighbourhood stars: $\Usun$ and $\Wsun$
are simply the negative radial and vertical components of these
means\footnote{According to the above definition, the LSR's radial and
  vertical motion w.r.t.\ the Galactic centre vanish. Therefore, the
  determination of $\Usun$ and $\Wsun$ from such means implicitly assumes that
  the Solar neighbourhood as a whole does not move radially or vertically
  w.r.t.\ the Galaxy. That such motions are at most small is suggested by the
  proper motion of Sgr\,A$^\star$ \citep{ReidBrunthaler2004} and the mean
  radial velocity of the stars orbiting it
  \citep[e.g.][]{ReidEtAl2007}. Moreover, such motions should also obey an
  asymmetric-drift like relation (see below), i.e.\ the mean velocities depend
  systematically on velocity dispersion, which is not observed.}.

The component $\Vsun$ of $\vsun$ in the direction of Galactic rotation is much
harder to determine, because the mean lag with respect to the LSR, the
asymmetric drift $\upsilon_{\mathrm a}$, depends on the velocity dispersion
$\sigma$ of the respective stellar population. The classical solution to this
problem exploits the empirical linear relation between the negative mean
heliocentric azimuthal velocity of any stellar sample
$\overline{\upsilon}_s=\upsilon_{\mathrm a}+\Vsun$ and its $\sigma^2$
\citep{Stromberg1946}. Hence, a straight-line fit yields $\Vsun$ as the value
of $\overline{\upsilon}_s$ for a hypothetical population of stars on circular
orbits, for which $\sigma=0$.

The theoretical underpinning of this method is the asymmetric drift relation
\citep[see][eq.~4.228]{BinneyTremaine2008}
\begin{equation}\label{eq:strom}
  \overline{\upsilon}_s-\Vsun =\va\simeq
  \frac{\overline{\upsilon^2_{\!R}}}{2\upsilon_{\mathrm{c}}}
  \left[\frac{\sigma^2_{\phi}}{\vrq} - 1 -
    \frac{\partial\ln(\nu\vrq)}{\partial\ln R} -
    \frac{R}{\vrq} \frac{\partial(\overline{\upsilon_R\upsilon_z})} {\partial z}
  \right],
\end{equation}   
where $R$ is Galactocentric cylindrical radius, $z$ the height above the
plane, $\upsilon_{\mathrm{c}}$ the circular speed, and $\nu$ the number
density of stars, while a bar indicates a $\nu$-weighted local mean. The
equation applies separately to each relaxed stellar population, for example to
M stars or G stars. The idea behind the classical determination of $\Vsun$ is
that the square bracket in equation (\ref{eq:strom}) takes essentially
identical values for each stellar population, with the consequence that a plot
of $\overline{\upsilon}_s$ against $\vrq$ should be linear.

\citeauthor{DehnenBinney1998:Local} (\citeyear{DehnenBinney1998:Local},
hereafter DB98) applied this method to a sample of ${\sim}\,15\,000$
main-sequence stars from the Hipparcos catalogue and their value of
$\Vsun=(5.25\pm0.62)\kms$ has been widely used. Recent re-determinations
using an improved reduction of the Hipparcos data \citep{vanLeeuwen2007}
confirm the DB98 value though with reduced error bars
\citep{vanLeeuwen2007,AumerBinney2009}.

However, two recent studies call the DB98 value for $\Vsun$ into question.
\citeauthor{Binney2009} (\citeyear{Binney2009}, hereafter B09) fitted
distribution-function models (a) to velocity distributions inferred by
\cite{IvezicEtAl2008} from proper motions and photometric distances of stars
in the Sloan Digital Sky Survey, and (b) to the space velocities of F and G in
the Geneva-Copenhagen Survey \citep[GCS,][]{NordstromEtal2004}. The GCS stars
are a subset of the Hipparcos stars (analysed by DB98) for which radial
velocities have been obtained. B09 was able to obtain satisfactory fits to
these data only if $\Vsun$ was larger than the DB98 value by $\sim6\kms$,
about ten times the formal error on $\Vsun$. Another body of evidence
against the DB98 value for $\Vsun$ originates from radio-frequency
astrometry of masers in regions of massive-star formation \citep{RyglEtAl2009,
ReidEtAl2009}. If the DB98 value for $\Vsun$ is correct, these sources
systematically lag circular rotation by $\sim17\kms$
\citep{ReidEtAl2009}. Such a high systematic lag is unexpected for young stars
and \cite{McMillanBinney2009} argued that a more plausible interpretation of
the data is obtained if $\Vsun$ exceeds the DB98 value by $\sim6\kms$.

This paper does two things: (i) it explains why the approach to the
determination of $\Vsun$ by DB98 and subsequent studies is misleading, and
(ii) it determines $\Vsun$ from similar data but a different methodology.
Both these tasks are accomplished with the help of a particular
chemo-dynamical model of the Galaxy, that of
\citeauthor{SchonrichBinney2009:Chemo} (\citeyear{SchonrichBinney2009:Chemo},
hereafter SB09a), but the points that we make are general ones and the role
played by the SB09 model is essentially illustrative. In
Section~\ref{sec:goveqs} we show that a metallicity gradient in the disc
gives rise to distributions of mean azimuthal velocity and velocity
dispersion within the colour-magnitude plane that are much more complex than
one naively expects, and we show that these distributions invalidate the
methodology of DB98.  In Section~\ref{sec:Vsun} we re-estimate $\Vsun$ by
fitting the entire velocity distribution of the GCS stars to the distribution
predicted by the SB09 model without reference to the Str\"omberg relation.

%%%%%%%%%%%%%%%%%%%%%%%%%%%%%%%%%%%%%%%%%%%%%%%%%%%%%%%%%%%%%%%%%%%%%%%%%%%%%%%%
\section{Kinematics in colour and magnitude}\label{sec:goveqs}

DB98 divided their sample of Hipparcos main-sequence stars into populations
with different velocity dispersions by binning in \BV\ colour because colour
is correlated with age and therefore with velocity dispersion.  To examine
the relation between colour, mean rotation velocity and velocity dispersion
for stars near the Sun, we employ the SB09a model of the Galactic Disc. This
model describes the chemodynamical evolution of the thin and thick Galactic
discs and is a refinement of models pioneered by \cite{vandenBergh1962} and
\cite{Schmidt1963}. The disc is divided into 80 annuli, within each of which
the chemical composition of the ISM evolves in response to the ejection of
material by dying stars, while stars form continuously with the current
composition of the ISM. The new features of the model are (a) stochastic
stellar accelerations accounting for heating processes; (b) radial stellar
migration accounting for both non-circular orbits and guiding-centre shifts
caused by stochastic resonant scattering off spiral arms
\citep{SellwoodBinney2002}; and (c) transfer of gas between annuli, both as
result of resonant scattering by spiral arms and as a result of a secular
tendency of gas to spiral inwards through the disc. Surprisingly, the model
contains both thin and thick discs that are consistent with the available
observational constraints \citep{SchonrichBinney2009:Disc}.

%-----------------------------------------------------------------------
\begin{figure}
  \epsfig{file=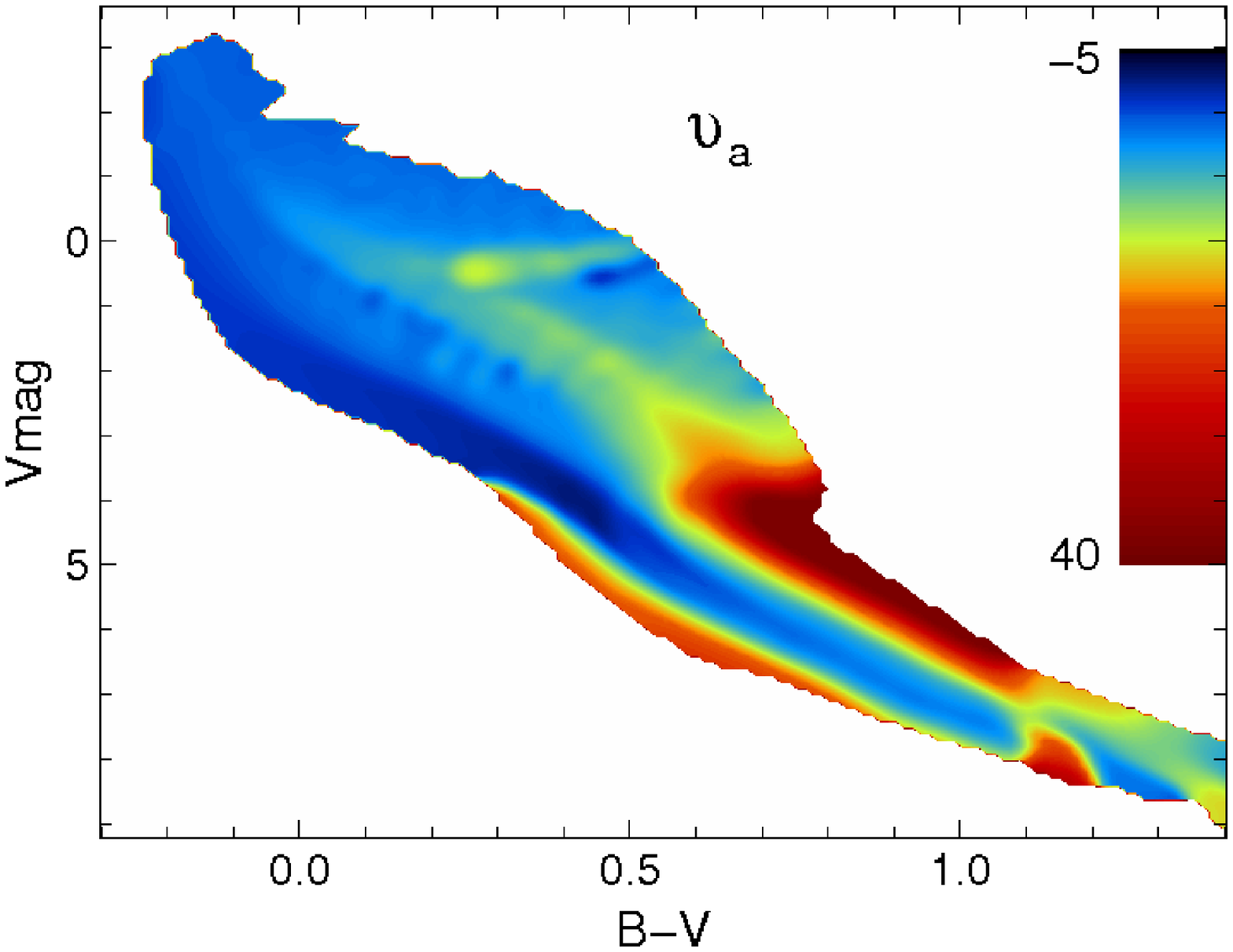,width=\hsize}
  \epsfig{file=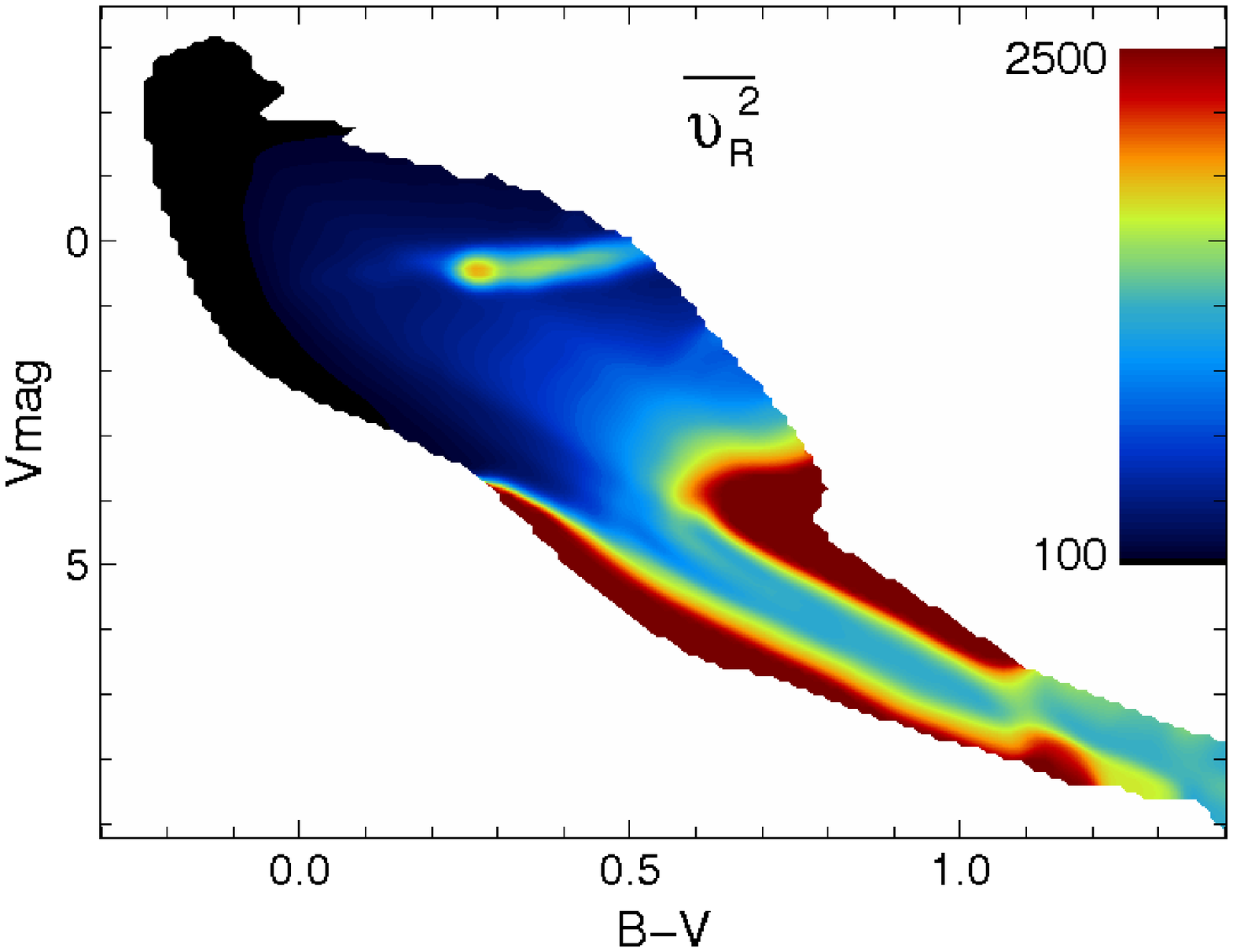,width=\hsize}
  \caption{The variation with colour and magnitude of the asymmetric drift
    $\va$ (top) and radial velocity dispersion (bottom) in the SB09a model of
    the Solar neighbourhood. Shown is the range of colours and magnitudes used
    by DB98 to generate their main-sequence sample. Note that the number
    density of stars is highly non-uniform across the region shown.}
  \label{fig:HRv}
\end{figure}
%-----------------------------------------------------------------------
\figref{fig:HRv} shows the model kinematics in the colour-magnitude diagram.
Each point in colour-magnitude space defines a separate sub-population whose
asymmetric drift $\va\equiv\upsilon_{\mathrm{c}}-\overline{\upsilon}_\varphi$
and radial velocity dispersion are plotted via colour coding, such that
dynamically cold and warm populations are shown with blue and  red shades,
respectively. The region in the colour-magnitude diagram shown in this figure
corresponds to the cuts used by DB98 to define their sample.

Our naive expectation is that as we proceed down the main sequence from its
blue end towards the main-sequence turnoff at $\BV\sim0.6$, we encounter
successively older stars with lower mean rotation velocities and higher
velocity dispersions, so in both panels of \figref{fig:HRv} the shading
should become redder as we move from left to right along the main sequence.
The pattern actually found in \figref{fig:HRv} is more complex. Most notably,
there is a pronounced velocity gradient \emph{across} the lower main sequence.
In the range $0.6<\BV<1$ the lower edge of the main sequence is dynamically
warm (orange in \figref{fig:HRv}) on account of subdwarfs, which are
metal-poor and therefore old with large velocity dispersions and low mean
rotation rates.  The number of these subdwarfs is small, however, so they
will not have a significant impact on a sample binned by colour alone. More
significant is the orange shading on the upper edge of the lower main
sequence, which reflects the metallicity gradient within the disc: as
metallicity increases, the main sequence shifts to the right, so in the upper
panel the orange upper edge of the main sequence implies that the more
metal-rich stars of the Solar neighbourhood are rotating more slowly because
they formed at $R<R_0$. To the left of $\BV\simeq0.5$ this trend is
weakened by contributions from old, sometimes metal-poor populations whose
isochrones move up through this region. Still the more metal-rich
main-sequence stars with smaller guiding centre radii give rise to
slightly higher dispersions and asymmetric drifts to the red side of
the main sequence.

The upper panel of \figref{fig:HRv} shows that in the crucial colour range
$0.4<\BV<0.6$, the asymmetric drift is a complex function of colour and
absolute magnitude because in this region stars of widely differing ages and
metallicities are found as a result of old, metal-poor isochrones intersecting
younger, metal-rich isochrones.

The  horizontal branch is clearly visible on both panels of \figref{fig:HRv} as
an almost horizontal feature just below $M_V=0$. It is less pronounced in the
upper panel because the blue end of the horizontal branch contains metal-poor
stars, which tend to have large guiding-centre radii and therefore low $\va$
even at large $\vrq$.

%-----------------------------------------------------------------------
\begin{figure}
  \epsfig{file=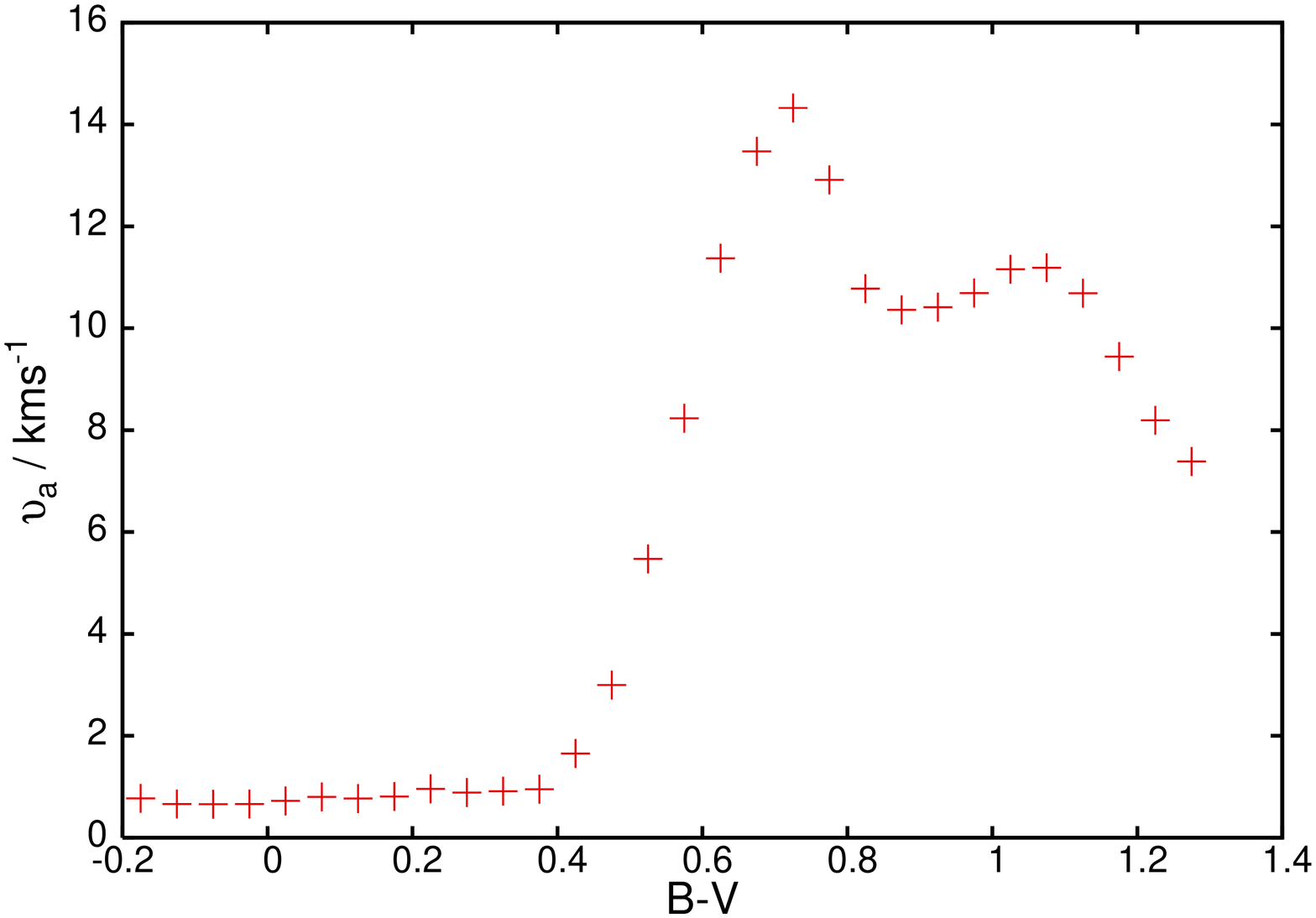,angle=-90,width=78mm}
  \epsfig{file=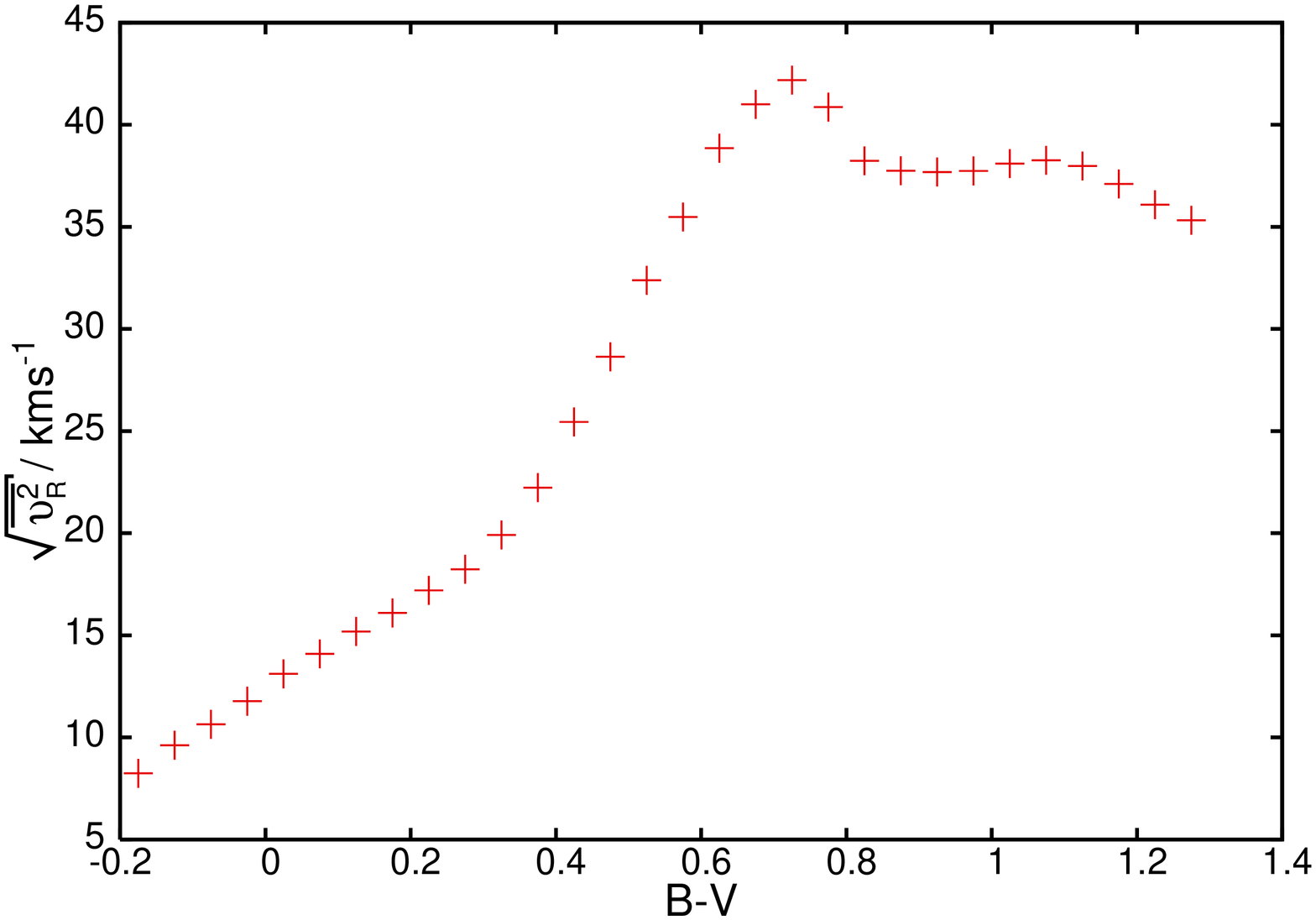,angle=-90,width=78mm}
  \caption{The asymmetric drift (top) and radial velocity dispersion (bottom)
    for stellar samples of given \BV\ colour drawn from the model thus
    simulating the effects of the Hipparcos and DB98 selection
    criteria.}\label{fig:chromosv}
\end{figure}
%-----------------------------------------------------------------------

\figref{fig:chromosv} shows the asymmetric drift $\va$ and velocity dispersion
$\vrq$ obtained when  stars are binned by colour alone. The lower panel can be
compared with corresponding observational plots in DB98 and
\cite{AumerBinney2009}. The model reproduces the structure of the data very
well -- in particular, the steepening in the slope around $\BV=0.4$ and the
flatness redwards of $\BV=0.6$.  The peak in velocity dispersion seen in the
lower panel of \figref{fig:chromosv} is much less evident in Fig.\,2 of
\cite{AumerBinney2009} but can be traced in their $\sigma_R$ and $\sigma_z$
data. Note that the rise with $\BV$ in $\vrq$ for $\BV<0.4$ is not accompanied
by any change in $\va$. This unexpected phenomenon arises because at these
colours the contribution of old metal-poor stars is increasing with $\BV$, and
because we see many of these stars near pericentre, they have small asymmetric
drifts despite their large random velocities.

Since at its bright end the Hipparcos sample is close to being volume
limited, the relative number of horizontal-branch stars is small and the
complex structure above the main sequence in \figref{fig:HRv} has no
significant effect in \figref{fig:chromosv}.  

%-----------------------------------------------------------------------
\begin{figure}
  \epsfig{file=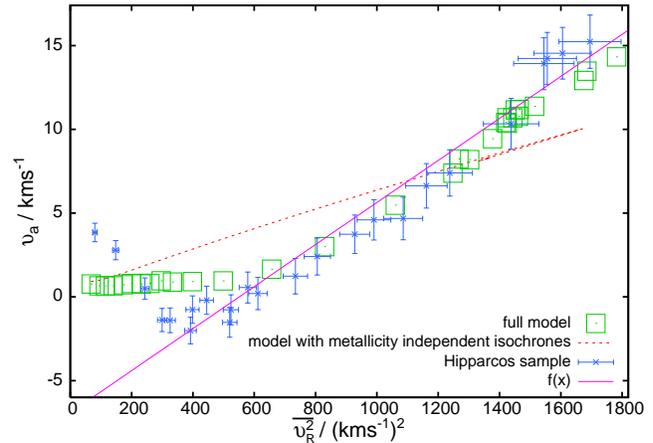,angle=-90,width=\hsize}
  \caption{Green squares: the asymmetric drift for synthetic stellar
  sub-samples defined by 
    \BV\ colour plotted against their radial velocity dispersion squared for
    the SB09a model. The red dots:  the same relation
    obtained from a model with only one chemical composition
    (solar). Blue data points: the values of
    Aumer \& Binney (2009) shifted by $11 \kms$ and with the radial
    velocity dispersion increased by $7 \%$. Purple line: a linear fit to the
    Hipparcos points in the range of $\vrq$ used by DB98.}\label{fig:chrom}
\end{figure} 
%-----------------------------------------------------------------------

In \figref{fig:chrom}, the squares show the resulting plot of the asymmetric
drift $\va$ against velocity dispersion squared for the synthetic samples of
\figref{fig:chromosv}; they do not lie on a straight line. The red dots
show the plot one obtains if all stars are assigned solar
metallicity. These dots \emph{do} lie on a good approximation to a straight
line.\footnote{The slight deviation from a straight line of the model without
  metallicities is an effect of the approximations in SB09a and leads to a
  small underestimation of the real metallicity bias by the model.}  The
effect of re-assigning stars with large guiding-centre radii from solar
metallicity to their true, low metallicities is to move them from redder to
bluer bins. Since these stars have small or even negative values of $\va$ on
account of the metallicity gradient in the disc, the transfer reduces $\va$
for the young, bluer bins and increases it in the old, redder
bins. Consequently, the transfer morphs the near-straight line of the red
points into the curve defined by the green squares.

DB98 estimated $\Vsun$ by fitting a straight line to the observational
analogue of \figref{fig:chrom}, which is a plot of the solar velocity
relative to a colour-selected group of stars versus the squared velocity
dispersion of that group. The blue data points in \figref{fig:chrom} show
such data for the Hipparcos sample in the re-analysis of
\cite{AumerBinney2009} after subtracting $11\kms$ from each value of solar
velocity.
We see that for $\vrq\ga600\,[\!\kms]^2$ the Hipparcos data define the same
straight line as the green crosses from the model. This straight line
intercepts the $\va$ axis at $\sim-7\kms$ rather than 0, causing $\Vsun$ to
be underestimated by this amount. For $\vrq\la400\,[\!\kms]^2$ the Hipparcos
data points in \figref{fig:chrom} deviate from this straight line, but DB98
ignored samples with very low velocity dispersion on the grounds that such
samples may not be in dynamical equilibrium. Indeed, both dissolving star
clusters and the non-axisymmetric gravitational potentials of spiral arms are
liable to distort the kinematics of stellar samples with low random velocities
such that equation~(\ref{eq:strom}) does not hold.

%-------------------------------------------------------------------------------
\begin{figure}
  \epsfig{file=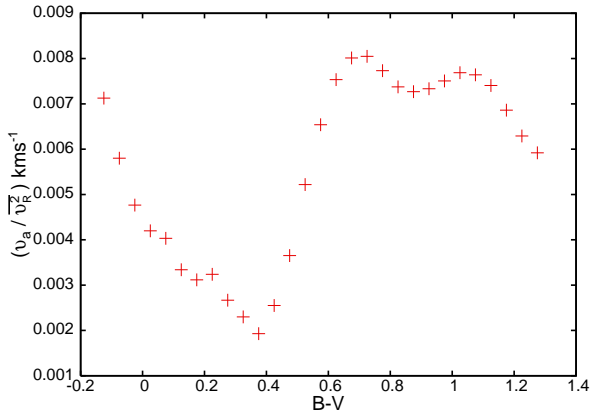,angle=-90,width=78mm}
  \caption{Asymmetric drift velocity divided by squared velocity dispersion
    for synthetic colour-selected samples using the SB09a model. This ratio
    should be proportional to the square bracket in equation
    (\ref{eq:strom}).}\label{fig:strombrack}
\end{figure}
%-------------------------------------------------------------------------------

The failure of the synthetic samples to follow a straight line in
\figref{fig:chrom} implies that the square bracket in the asymmetric drift
relation~(\ref{eq:strom}) does depend on colour: it varies by a factor $\sim4$
in the colour range $0.4<\BV<0.6$ as demonstrated in \figref{fig:strombrack}.
A significant contribution to the value of
the bracket comes from the first derivative term, which 
is smallest for metal-poor populations because their densities
$\nu$ decline more slowly outwards on account of the metallicity gradient in
the disc.  Physically, including metal-poor, thin-disc stars decreases $\va$
because such stars typically visit the Solar neighbourhood at pericentre,
where they have $\upsilon_\phi>\vc$.

%%%%%%%%%%%%%%%%%%%%%%%%%%%%%%%%%%%%%%%%%%%%%%%%%%%%%%%%%%%%%%%%%%%%%%%%%%%%%%%%
\section{determining the Solar motion from the Velocity distribution}
\label{sec:Vsun}
In view of the argument just presented, that the classical procedure cannot
yield a reliable value of $\vsun$, we now estimate $\vsun$ by fitting the observed
distributions of heliocentric velocities to the velocity distribution of the
SB09a model. That is, we seek the offset $-\vsun$ from the circular velocity
at which the model velocity distributions provide the best match to the
distribution of observed heliocentric velocities.  B09 used an analogous
procedure to argue that $\Vsun\simeq11\kms$; however, the model distributions
he used were obtained from analytic distribution functions rather from a model
of the Galaxy's chemical evolution.  By using velocity distributions that
reflect much prior information about the chemodynamical history of the Galaxy
in place of simple analytic functions, we hope to achieve a closer fit to
the observed velocity distributions and therefore determine the requisite
offset $-\vsun$ with greater precision.

%-------------------------------------------------------------------------------
\begin{figure}
  \epsfig{file=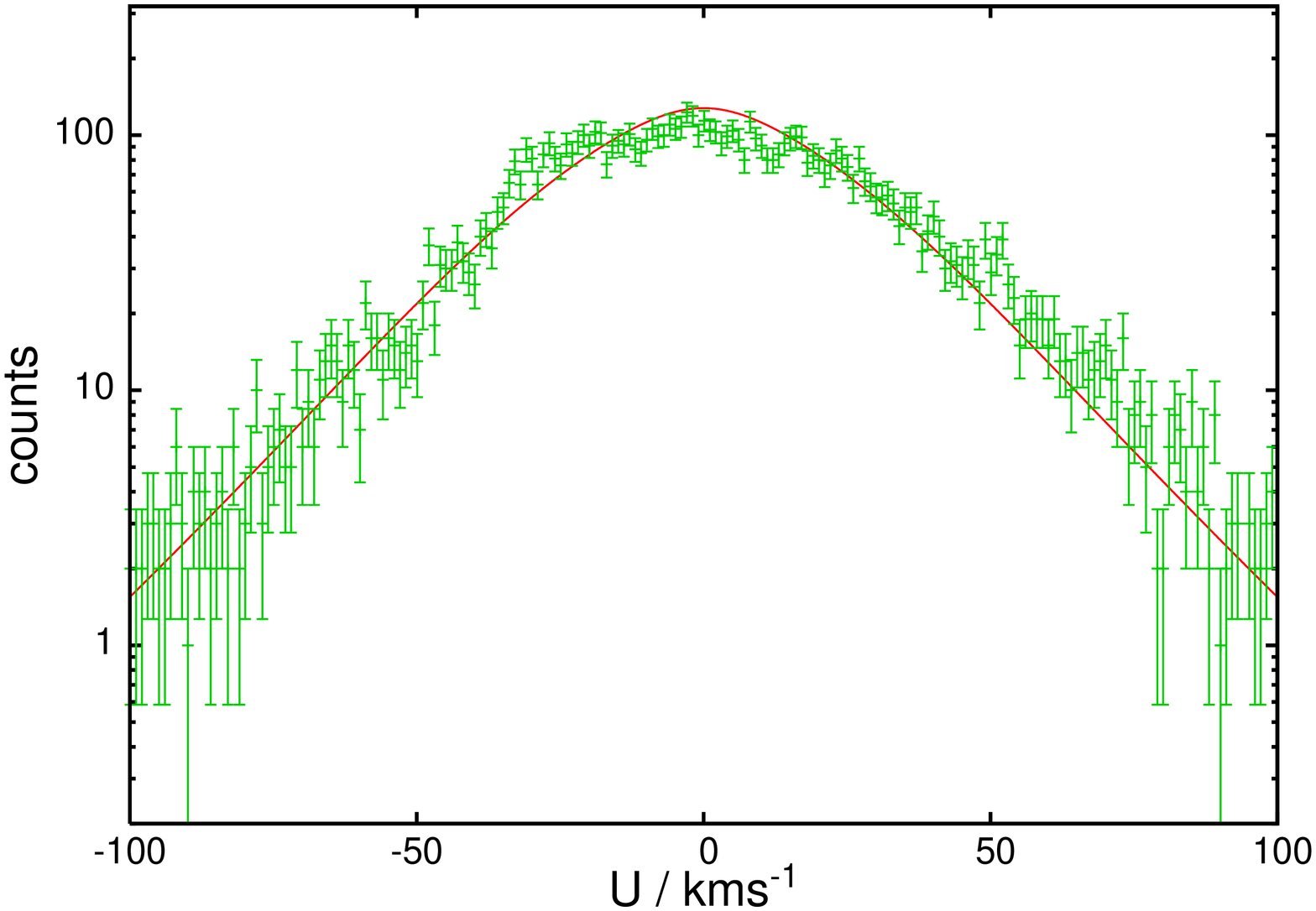,angle=-90,width=77.9mm}
  \epsfig{file=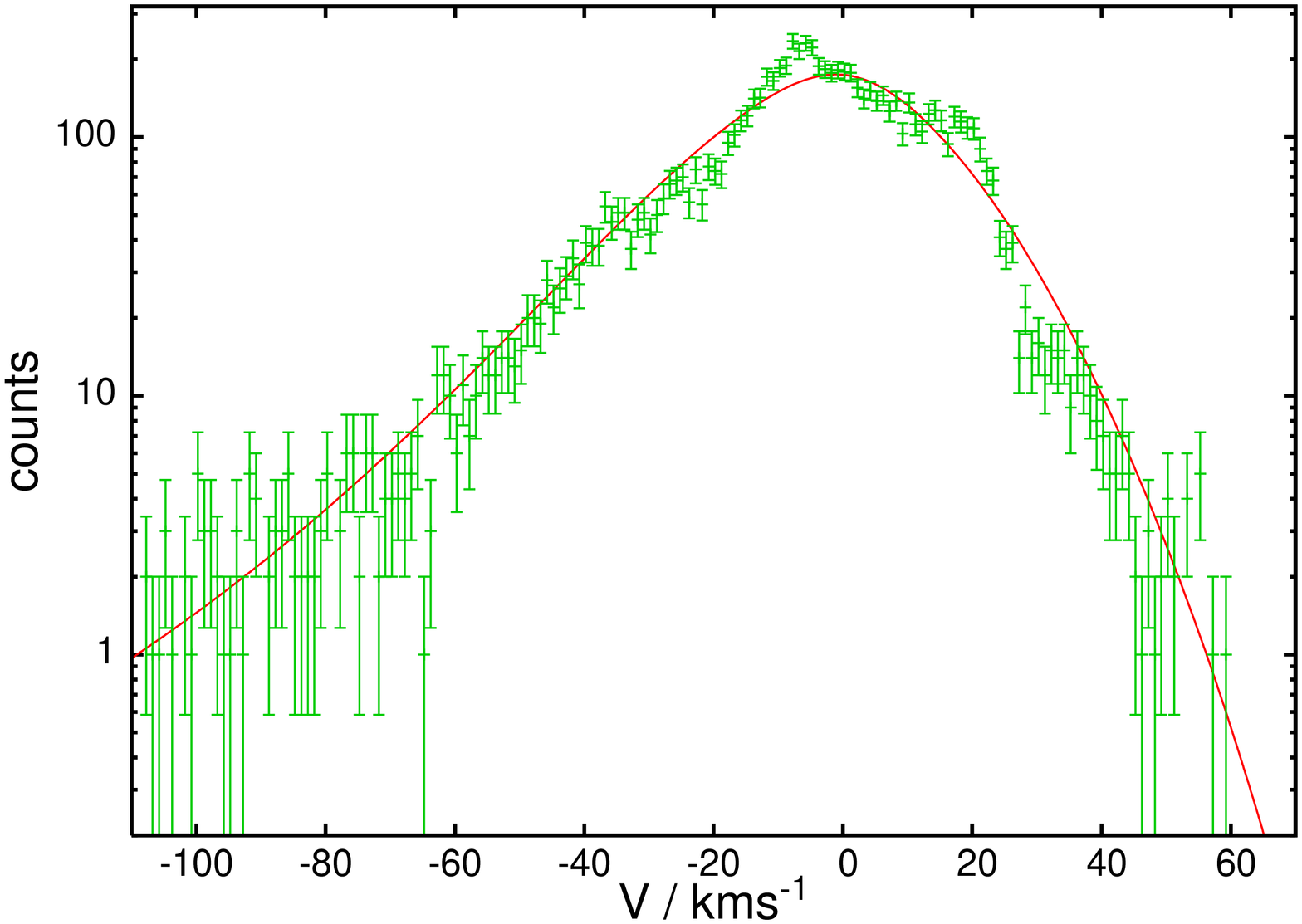,angle=-90,width=77.9mm}
  \epsfig{file=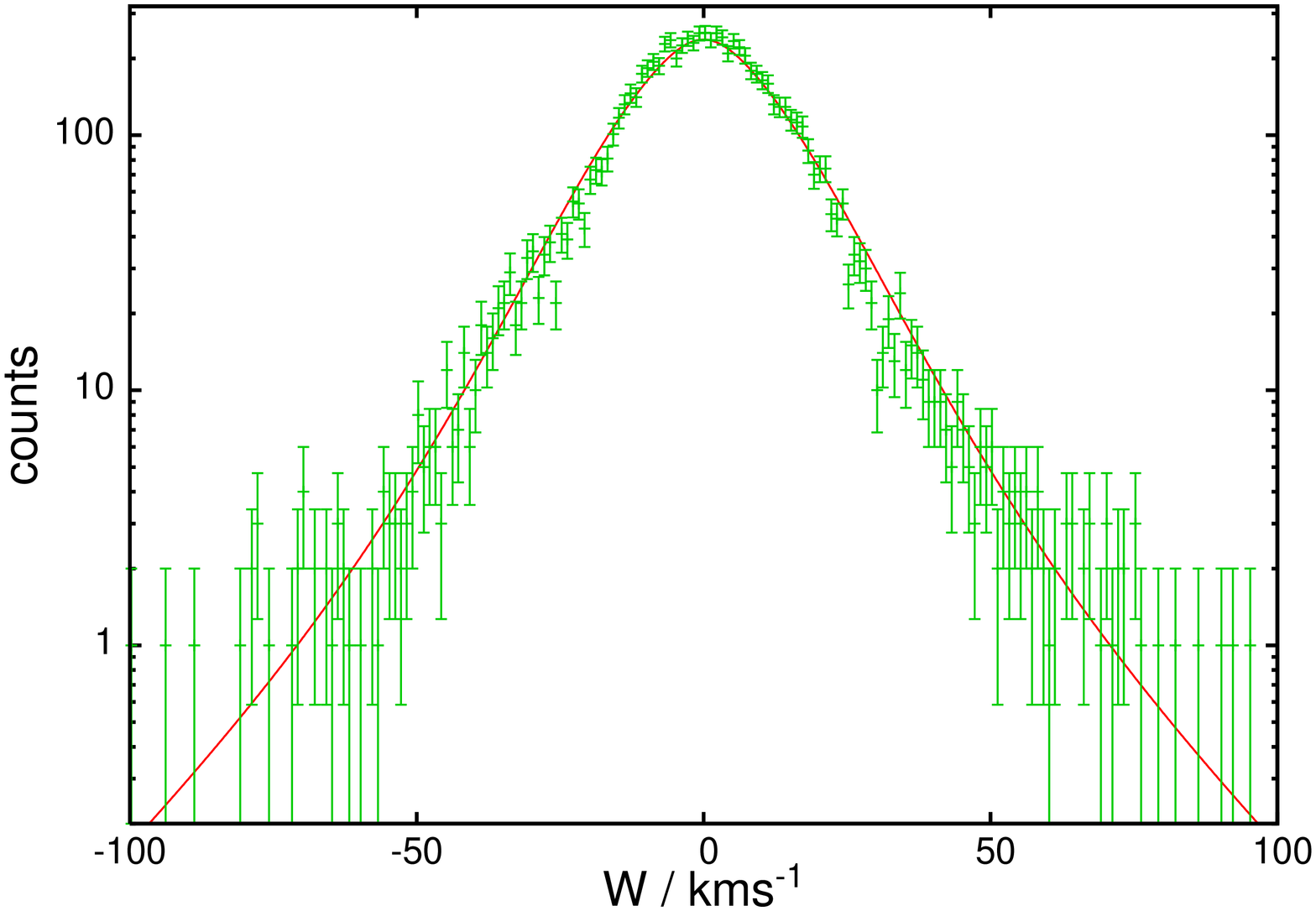,angle=-90,width=77.9mm}
  \caption{Curves: The model distributions predicted by the
    SB09a model in the $U$, $V$ and $W$ components of velocity (from
    top to bottom). Data points with Poisson error bars: the observed
    distributions of the GCS stars shifted by our estimate of $\vsun$
    to optimise the fit of the data.}\label{fig:veldis}
\end{figure}
%-------------------------------------------------------------------------------

In \figref{fig:veldis} the points with (Poisson) error bars are for a
subsample of GCS stars for which
\cite*{HolmbergNordstromAndersen2007,HolmbergNordstromAndersen2009} give
reliable metallicities; thirty likely halo stars have been removed by
requiring $[\mathrm{Fe/H}]>-1.2$. This criterion is slightly
stricter than that used in SB09b for the determination of the
in-plane dispersion parameter ($\sigma_R$ of a $10 \Gyr$ old local 
population), such that we now use a marginally smaller
value, $43\kms$, while lowering the vertical dispersion parameter to
$23\kms$. The curves in \figref{fig:veldis} show the 
model distributions when offset by
$-\vsun=(11.10_{-0.75}^{+0.69},12.24_{-0.47}^{+0.47},7.25_{-0.36}^{+0.37})\kms$
-- we used cubic splines to interpolate between individual data points
provided by the model in the $V$ component and determined the offset by
maximising the likelihood of the data given the model.
We used only five parameters for all three distributions: the two dispersion parameters and the three
components of Solar motion, yet the fit is of good quality.
The small fluctuations of the
data around the model $V$ and to a lesser extent $U$ distributions are readily
accounted for by the well known stellar moving groups \citep{Dehnen1998},
likely caused by the dynamical influence of the Galactic bar and spiral
structure
\citep{Dehnen1999:Bar,DeSimoneWuTremaine2004,AntojaEtAl2009,MinchevEtAl2009}.

A significant advantage of determining $\vsun$ from the entire sample, as in
this section, rather than from subsamples as done in the past, is the
robustness of the result to changes in the modelled metallicity gradient.
In fact, eliminating the model's metallicity gradient changes $\Vsun$ by less than
$0.1\kms$.

One should note that the quoted errors on the components of $\vsun$ are
purely formal. Sources of additional systematic error include the possible
presence of halo stars in the sample, the dynamical approximations used in
constructing the model, and the effects of stellar streams, which have a big
impact on the observed distribution of stars near the circular velocity but
are completely excluded from the model. Fortunately the likelihood of the
data used here is not particularly sensitive to the fit of the data to the
model around the peak density.  In view of these uncertainties we roughly
estimate systematic errors of $\sim(1,2,0.5)\kms$, assuming that $U$ and $W$
are mostly affected by distortions by streams and $V$ showing even more
structure and having an additional uncertainty from the modelling.  This is
in perfect agreement with previous estimates as regards $\Wsun$ and slightly
higher in $\Usun$ compared to the DB98 value, which can be traced back of the
larger influence of the Hercules stream at $\sim -30\kms$ (lowering the
estimate) on their statistics. However, it differs significantly from the
value for $\Vsun\simeq5.2\pm0.5\kms$ obtained by the classical technique
\citep[DB98,][]{AumerBinney2009}. Our value for $\Vsun$ is in good agreement
with $\Vsun\simeq11\kms$ proposed by {B09}. Given the residual uncertainties
of $\vsun$, it is questionable whether the standard practice of
``correcting'' observed (heliocentric) velocities for the Solar motion is
useful, at least the adopted value should be explicitly provided.
  
%%%%%%%%%%%%%%%%%%%%%%%%%%%%%%%%%%%%%%%%%%%%%%%%%%%%%%%%%%%%%%%%%%%%%%%%%%%%%%%%
\section{Conclusions} \label{sec:disc}
The metallicity gradient in the Galactic disc causes a systematic
shift in the kinematics especially near the turnoff region. By the
relationship between the colour and metallicity of a star, the more 
metal-rich populations, with on average lower angular momentum and thus
higher asymmetric drifts, are displaced relative to their metal-poor
counterparts, which have lower asymmetric drifts.  
When stars are binned by colour, the metallicity gradient
in the Galactic disc prevents the relationship between mean rotation speed
and squared velocity dispersion taking the linear form predicted by a naive
application of the Str\"omberg relation. This breakdown in the conventionally
assumed linearity invalidates the traditional technique for determining the
Sun's velocity with respect to the LSR, which involves a linear extrapolation
to zero velocity dispersion of the empirical relation between the mean
velocity and squared velocity dispersion. Moreover the SB09a model
predicts that a treacherous linear relationship underestimating the
solar azimuthal motion by $\sim 7\kms$ will be mimicked redwards of
the onset of the turnoff region, coinciding well with the behaviour
observed in the Hipparcos data.  

The Sun's velocity with respect to the LSR may be alternatively determined
from the velocity offset that optimises a model fit to the observed velocity
distribution. Using the velocity distribution predicted by the SB09a model of
the chemodynamical evolution of the Galaxy, we find
$\vsun=(11.1_{-0.75}^{+0.69}, 12.24_{-0.47}^{+0.47},
7.25_{-0.36}^{+0.37})\kms$ and roughly estimate the systematic uncertainties
as $\sim(1,2,0.5)\kms$. The radial and vertical components of this value of
$\vsun$ agree with earlier estimates, but the $V$ component is larger than the
widely used value of DB98 by $\sim7\kms$. This is in nice concordance with the
model expectations for the systematic error arising from naively using the
Str\"omberg relation and in good agreement with the value obtained by {B09}
using a similar method but with a less sophisticated distribution function.
Curiously it agrees well with the result $\Vsun \sim 10-13 \kms$
obtained by \cite{Delhaye1965} using the classical method with pre-Hipparcos data.

In this paper we have relied heavily on the SB09a model, so the question
arises of how vulnerable our argument is to the model's shortcomings. Our
critique of the classical approach to the determination of $\vsun$ is secure
so long as the disc has significant age and/or metallicity gradients. It is
beyond question that such gradients exist, so the classical technique is
certainly unreliable. Our proposed value of $\Vsun$ is essentially
independent of the assumed metallicity gradient, but does have some
sensitivity to the dynamical approximations used in making the SB09a
model -- plausible variations of how one handles the secular
acceleration of stars lead to changes in the estimated value of $\Vsun$ by $0.5$ to $1\kms$.
Modest reassurance that the error of our value of $\Vsun$ is less than
$2\kms$ is furnished by the fact that B09 favoured the same value using a
distribution function that takes no account of the age and metallicity
gradients in the disc. Consequently, our result is probably not sensitive to
the assumptions about star formation and chemical evolution made by SB09a.
However, as we develop more elaborate models of the Galaxy which fit a wider
range of data, in particular more distant stars, we anticipate
further small revisions in the value of $\Vsun$.

\section*{Acknowledgements}
We thank Michael Aumer for kindly providing and discussing his data, and
Martin Asplund and Paul McMillan for valuable comments on an early
draft. R.S.\ acknowledges financial and material support from
Max-Planck-Gesellschaft and Max Planck Institute for Astrophysics.

%%%%%%%%%%%%%%%%%%%%%%%%%%%%%%%%%%%%%%%%%%%%%%%%%%%%%%%%%%%%%%%%%%%%%%%%%%%%%%%%
%
% Use the following with bibtex to automatically generate the .bbl file and
% the reference list. You do need the files refs.bib and mn2e.bst.
% If you use this (not commented out), you don't need to comment out the stuff
% after "\end{document}", of course.
%

%\bibliographystyle{mn2e}
%\bibliography{refs}
%\label{lastpage}
%\end{document}
%%%%%%%%%%%%%%%%%%%%%%%%%%%%%%%%%%%%%%%%%%%%%%%%%%%%%%%%%%%%%%%%%%%%%%%%%%%%%%%%
%
% Code below is only seen if the four lines above are commented out. 
%
% The following is just the .bbl file generated via bibtex copied here and
% modified by adding "(DB98)", "(B09)", and "(SB09a)" and other minor edits to
% correct for errors in mn2e.bst (reference to Ivezic et al).
%

\label{lastpage}
\end{document}